\documentclass[journal,twoside,web]{ieeecolor}
\usepackage{tmi}
\usepackage{amsmath,amssymb,amsfonts}
\usepackage{algorithm}
\usepackage{algorithmic}
\usepackage{graphicx}
\usepackage{textcomp}
\usepackage{multirow}
\usepackage{booktabs}
\usepackage[table]{xcolor} 
\usepackage{colortbl}      
\usepackage[colorlinks,
            linkcolor=blue,       
            anchorcolor=red,  
            citecolor=black,        
            ]{hyperref}
\usepackage{diagbox}
\renewcommand{\algorithmicrequire}{\textbf{Input:}}  
\renewcommand{\algorithmicensure}{\textbf{Output:}} 
\usepackage{ragged2e}
\usepackage{verbatim}

\usepackage{mdframed}
\newmdenv[backgroundcolor=yellow, linewidth=0pt]{highlighted}

\usepackage{rotating}
\usepackage{cleveref}

\usepackage[backend=biber, style=ieee]{biblatex}  
\usepackage{soul}
\sethlcolor{yellow}
\soulregister\cite7 
\soulregister\ref7

\crefformat{section}{Section~#2#1#3} 
\crefformat{subsection}{Section~#2#1#3} 
\crefformat{subsubsection}{Section~#2#1#3} 
\crefformat{figure}{Figure~#2#1#3} 

\crefname{section}{Section}{Sections}
\crefname{subsection}{Section}{Sections}
\crefname{subsubsection}{Section}{Sections}
\crefname{figure}{Figure}{Figures}

\def\BibTeX{{\rm B\kern-.05em{\sc i\kern-.025em b}\kern-.08em
    T\kern-.1667em\lower.7ex\hbox{E}\kern-.125emX}}
\begin{document}
\title{Spatial and Modal Optimal Transport for Fast Cross-Modal MRI Reconstruction}
\author{Qi Wang, Zhijie Wen, Jun Shi, Qian Wang, Dinggang Shen, \IEEEmembership{Fellow, IEEE}, and Shihui Ying, \IEEEmembership{Member, IEEE}
\thanks{This work was supported by the National Key R \& D Program of China under Grant 2021YFA1003004 and the National Natural Science Foundation of China under Grant 11971296. (Corresponding author: Shihui Ying. (e-mail: shying@shu.edu.cn))}
\thanks{Qi Wang, Z. Wen, and S. Ying are with the Department of Mathematics, School of Science, Shanghai University, Shanghai 200444, China. }
\thanks{J. Shi is with the Key laboratory of Specialty Fiber Optics and Optical Access Networks, Joint International Research Laboratory of Specialty Fiber Optics and Advanced Communication, Shanghai Institute for Advanced Communication and Data Science, School of Communication and Information Engineering, Shanghai University, Shanghai 200444, China}
\thanks{Qian Wang and D. Shen are with the School of Biomedical Engineering, ShanghaiTech University, Shanghai 201210, China, and also with Shanghai United Imaging Intelligence Co., Ltd., Shanghai 200030, China.}}

\maketitle

\begin{abstract}
\justifying{Multi-modal magnetic resonance imaging (MRI) plays a crucial role in comprehensive disease diagnosis in clinical medicine. However, acquiring certain modalities, such as T2-weighted images (T2WIs), is time-consuming and prone to be with motion artifacts. It negatively impacts subsequent multi-modal image analysis. To address this issue, we propose an end-to-end deep learning framework that utilizes T1-weighted images (T1WIs) as auxiliary modalities to expedite T2WIs' acquisitions. While image pre-processing is capable of mitigating misalignment, improper parameter selection leads to adverse pre-processing effects, requiring iterative experimentation and adjustment. To overcome this shortage, we employ Optimal Transport (OT) to synthesize T2WIs by aligning T1WIs and performing cross-modal synthesis, effectively mitigating spatial misalignment effects. Furthermore, we adopt an alternating iteration framework between the reconstruction task and the cross-modal synthesis task to optimize the final results. Then, we prove that the reconstructed T2WIs and the synthetic T2WIs become closer on the T2 image manifold with iterations increasing, and further illustrate that the improved reconstruction result enhances the synthesis process, whereas the enhanced synthesis result improves the reconstruction process. Finally, experimental results from FastMRI and internal datasets confirm the effectiveness of our method, demonstrating significant improvements in image reconstruction quality even at low sampling rates.}
\end{abstract}


\section{Introduction}
\label{sec:introduction}

\IEEEPARstart{M}{AGNETIC} resonance imaging (MRI) is widely used for disease diagnosis due to its clear and high-contrast images, absence of ionizing radiation, and ability to reveal internal pathological changes for treatment guidance \cite{77}. It has become a crucial technology in clinical and research settings. Multi-modal MRI has gained popularity in various applications, providing comprehensive information for AI-based disease diagnosis and overcoming limitations of single-modality analysis \cite{78}. For example, T1-weighted images (T1WIs) capture morphological and structural details, while T2-weighted images reveal edema and inflammation in the same region of the body. However, some modalities, such as T2-weighted images (T2WIs), require longer scan times due to extended repetition (TR) and echo times (TE). Prolonged image acquisition increases the risk of motion artifacts, especially in chest and abdominal MRI, which can significantly impact subsequent multi-modal image analysis. To address these challenges, researchers have focused on reconstructing MRI from under-sampled data by reducing the sampling in the $k$-space \cite{3,4,5,6,71,80,127}. For instance, Zhou \emph{et al.} introduced a dual-domain self-supervised method aimed at accelerating non-cartesian MRI reconstruction, which effectively preserves $k$-space self-similarity \cite{127}. However, these approaches solely rely on single-modal information, limiting its effectiveness. 

\begin{figure}[!htbp]
\centerline{\includegraphics[width=\columnwidth]{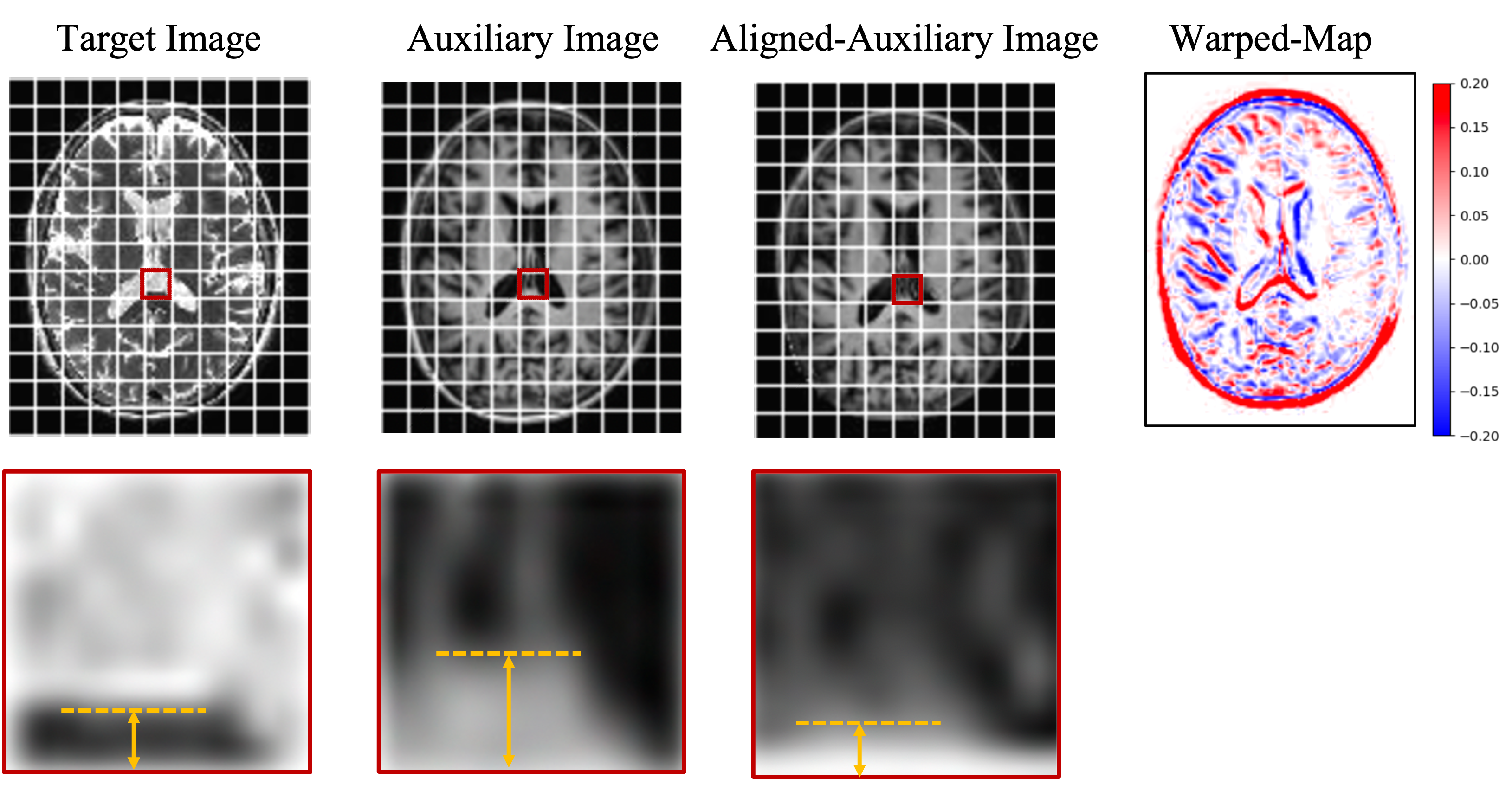}}
\caption{
In the zoomed-in views, we observe a slight spatial misalignment between the auxiliary (T1-weighted) image and the target (T2-weighted) image. The paper \cite{8} addresses this issue by using a spatial alignment method to generate the auxiliary alignment. The warped-map illustrates the disparity between the original and warped auxiliary (T1-weighted) images. The color transition from blue to red indicates an increasing difference between the two images.
}
\label{fig1}
\end{figure}

Consequently, the field has seen a growing interest in multi-modal MRI reconstruction \cite{102, 109, 126, 128, 129}. Liu \emph{et al.} designed a mechanism for shareable feature aggregation and selection among multi-modal MRI images, significantly enhancing the quality of the reconstructed images \cite{129}. Additionally, Li \emph{et al.} introduced a multi-scale transformer network with edge-aware pre-training as a solution for cross-modal magnetic resonance image synthesis \cite{109}. Their findings demonstrated that under-sampled target magnetic resonance images can be better reconstructed with the assistance of auxiliary modalities, which provide complementary information about the same region of interest. However, a common issue arises from the spatial differences prevalent between images from different modalities, even within the same subject. Fig. \ref{fig1} illustrates this spatial discrepancy between auxiliary images and target images, where the length of the yellow double-arrowed line in the target image is noticeably shorter than in the auxiliary image. This mismatch leads to negative information transfer. To address this challenge, various image pre-processing techniques have been proposed \cite{120, 121, 122, 123}. However, improper parameter selection during image pre-processing undermines the alleviation of misalignment, leading to undesired pre-processing outcomes. Therefore, iterative experimentation and adjustment are essential to address these issues effectively. From this viewpoint, Xuan \emph{et al.} proposed a novel deep learning method for image registration \cite{8}. Their approach involved the integration of a cross-modal registration module into a spatial alignment network during the reconstruction process. By leveraging the power of deep learning, their method demonstrated better results in achieving accurate and reliable image registration, as depicted in Fig. \ref{fig1}.

Although their approach achieves good reconstruction performance, using cross-modality-synthesis-based registration to indirectly measure similarity between two modality images is a big challenge due to substantial differences between modalities. Moreover, aligning auxiliary and under-sampled target images is problematic, especially at low sampling rates. For instance, at a $6.25\%$ undersampling rate, target images often lack crucial structural details. The absence of these essential features can substantially degrade the accuracy of the registration, undermining their method's utility in practical scenarios. Addressing these limitations is crucial for reliable and accurate registration.

Fortunately, the optimal transport (OT) framework provides a metric known as the Wasserstein distance, which measures the dissimilarity between distributions. The Wasserstein distance is often preferred over other divergence measures, such as the Kullback-Leibler divergence, in various applications like generative modeling \cite{111}. This preference arises from the topological and statistical properties exhibited by the Wasserstein distance. By utilizing the Wasserstein distance, researchers overcome the limitations associated with alternative divergence measures and achieve more desirable outcomes in tasks such as distribution alignment and modeling. Moreover, OT offers a powerful framework for interpreting the network by minimizing the Wasserstein distance between distributions directly \cite{82}. In their paper, the theoretical perspective of OT in Generative Adversarial Networks (GANs) brings transparency to a previously opaque process, transforming probability distribution through convex optimization based on OT theory. Based on this perspective, we utilize OT to enhance the spatial alignment process, facilitating coherent and meaningful information transfer across different modalities.

In this paper, we propose a comprehensive approach that integrates a reconstruction task and a synthetic task, with the objective of minimizing the discrepancy between their outputs on the T2 image manifold. To enhance the interpretability of the synthetic task, we introduce an OT-based network. Notably, our methodology involves decomposing the synthetic process into two key components: spatial alignment and cross-modal synthesis. In this way, we can better analyze these two different and relatively independent mappings.

By incorporating the reconstruction task and synthetic task, our approach offers a more interconnected framework that leverages the benefits of each task. The reconstruction task helps to improve the fidelity and accuracy of the generated images, while the synthetic task focuses on bridging the gap between different modalities and ensuring consistency. The utilization of an OT-based network further enhances the interpretability of the synthetic task, enabling a better understanding of the alignment process and the synthesis process.

By decomposing the synthetic process into spatial alignment and cross-modal synthesis, we address the challenges associated with modality differences and ensure a more reliable spatial alignment outcome. This approach not only improves the interpretability of the synthetic task but also enhances the overall quality and accuracy of the registration results.

In summary, the proposed framework makes the following main contributions:

\begin{itemize}
\item We introduce a novel synthetic process that combines spatial alignment and cross-modal synthesis within an OT framework. This approach leverages the spatial alignment mapping on the T1 image manifold, directly enhancing the alignment accuracy, while also considering the cross-modal synthesis mapping to bridge the gap between modalities.

\item We introduce an alternating iteration framework that integrates the reconstruction task and the synthetic task. Unlike traditional pre-processing approaches that treat these tasks independently, our framework demonstrates that these tasks are complementary and mutually reinforcing. This iterative process helps minimize the discrepancy between their outputs on the T2 image manifold, leading to improved overall performance.

\item The quantitative and qualitative experimental results prove that our model achieves better reconstruction results than state-of-the-art approaches on both an open dataset FastMRI and an in-house dataset.
\end{itemize}

\section{Related Work}
\label{sec:related work}

In this Section, Section \ref{DL} describes the development of MRI reconstruction and the reason for better performance of cross-modal reconstruction over single-modality reconstruction, and then current state of research on cross-modal image synthesis is discussed in Section \ref{IDA}. Finally, Section \ref{OTT} displays the superiority of the OT-based Learning methods.

\subsection{Deep Learning for Accelerating MRI}
\label{DL}

Accelerating MRI primarily focuses on recovering under-sampled $k$-space signals. To address this challenge, simple methods like zero-filling and linear filtering have been used, but they often introduce artifacts and compromise image quality. Based on the objective of accelerating MRI and addressing the challenges of under-sampled MRI reconstruction, a significant advancement has been achieved through the application of Compressed Sensing (CS) \cite{12}. CS enables the production of high-quality images sampled at much lower rates than the Nyquist sampling rate, given certain conditions on the sampling matrix. However, CS methods often rely on prior knowledge or assumptions about the sparsity of the signal, which limits their applicability. In recent years, deep learning-based methods have surpassed conventional techniques, establishing a new state-of-the-art in MRI reconstruction. Notably, Wang \emph{et al.} designed an offline prior algorithm that utilizes Convolutional Neural Networks (CNNs) to capture the relationship between zero-filled magnetic resonance images and ground-truth images \cite{17}. Additionally, subsequent studies have explored the advantages of data consistency layers \cite{26}, recurrent neural networks \cite{25}, dictionary learning \cite{27}, and federated learning \cite{1}, further pushing the boundaries of MRI reconstruction methods. For instance, Feng \emph{et al.} proposed a Federated Learning algorithm aiming at preserving useful details for local MRI reconstruction from under-sampled $k$-space data \cite{1}. While some methods overlook the spatial frequency properties and fail to utilize complementary information from adjacent slices, Du \emph{et al.} introduced adaptive CNNs for $k$-space data interpolation address the limitations \cite{24}.

In contrast to the aforementioned methods that independently reconstruct target images from a single MRI modality, this paper explores the use of an image from another modality to restore the target modality. Xiang \emph{et al.} initially employed a Dense-Unet architecture that takes data from different MRI modalities as input to accelerate the reconstruction of a target image \cite{6}. Moreover, Liu \emph{et al.} showcased that incorporating multi-modal MRI input into the reconstruction network enhances the anatomical accuracy of the reconstructed images \cite{126}. Based on the observation that $k$-space and image restoration are complementary, Zhou \emph{et al.} proposed DuDoRNet with T1 priors, which aims to recover both source modal images and target modal images \cite{3}. Recognizing the limitation of CNNs in capturing global knowledge, Feng \emph{et al.} introduced transformers into cross-modal reconstruction, incorporating a novel cross-attention mechanism for efficient fusion of different modalities \cite{4}. 
Additionally, Zhou \emph{et al.} combined hybrid operations that include both convolutional layers and Swin Transformer blocks to explore the relative information between multi-modal MRI images \cite{128}. This approach attained performance comparable to supervised reconstruction methods. However, most of these methods overlook the common issue of slight misalignment between paired images from different modalities. Inspired by Xuan \emph{et al.} \cite{8}, this work decomposes the cross-modal image synthesis process into two parts: spatial alignment and cross-modal synthesis, with the goal of enhancing the reconstruction quality.

\subsection{Cross-modal Image Synthesis}
\label{IDA}

Cross-modal image synthesis has emerged as a prominent area of research in computer vision, encompassing tasks such as super-resolution \cite{95}, style transfer \cite{92}, and image reconstruction \cite{8}. Traditionally, cross-modal image synthesis has been approached as a regression problem, where the input is an image in the source modality and the output is its corresponding image in the target modality. For instance, Jog \emph{et al.} employed a regression forest to establish nonlinear mappings between tissue modalities. However, the main challenge lies in learning an effective mapping from a source modality image to its target modality counterpart \cite{96}. In recent years, deep learning techniques utilizing CNNs have shown remarkable progress in expanding the mapping capabilities. Notably, Gulrajani \emph{et al.} proposed a PixelCNN-based model with an auto-regressive decoder to capture global features in generated images \cite{97}. Dong \emph{et al.} introduced a cross-domain adaptive filters module, known as MMSR, to learn the joint probability distribution of low-resolution images and high-resolution images \cite{95}.

OT \cite{15} provides a mathematical framework for measuring the similarity or dissimilarity between probability distributions, which can be applied to find the most efficient mapping between two distributions. In the context of cross-modal image synthesis, the goal is to generate target modality images that closely resemble the source modality while minimizing the OT cost. OT can be utilized to establish a mapping between the source modality and the target modality, aligning their respective distributions or manifolds. By leveraging this alignment, cross-modal image synthesis can generate more accurate and coherent synthetic images, overcoming the challenges of spatial misalignment and capturing the underlying information shared between modalities. For example, Tian \emph{et al.} leveraged OT to establish the mapping from text cues to generated images, effectively addressing the mode collapse issue and ensuring image diversity compared to simple metrics \cite{100}. Taking cues from the favorable outcomes resulting, we utilize OT to facilitate the synthesis process, ensuring consistency and meaningful transformation of information across different modalities.

\subsection{OT-based Learning Methods}
\label{OTT}

OT is a unique optimization problem that defines the Wasserstein distance between two probability distributions within a Lagrangian framework \cite{15, 62}. This distance can be estimated solely based on the empirical distribution of data, while maintaining a favorable convergence property even when the supports of the two distributions have a trivial intersection \cite{85}. Leveraging these advantageous characteristics, OT-based deep learning methods have found applications in diverse domains, including image reconstruction \cite{90}, generative models \cite{86}, and domain adaptation \cite{87}. OT empowers the exploration of optimal correspondences between disparate data modalities, facilitating the seamless transfer of significant and coherent information. For instance, OT is used as a mathematical framework to derive the architecture of the OT-cycleGAN for accelerated MRI reconstruction in this paper\cite{22}. It helps guide the mapping between the under-sampled $k$-space data and the fully-sampled image domain, facilitating the generation of high-resolution magnetic resonance images from accelerated data. Moreover, deep learning based methods have demonstrated remarkable performance in various tasks compared to traditional approaches, but their interpretability is limited. OT provides a topological interpretation of CNNs and offers insights into the learned data distribution. In the context of generative models like GANs, the main objectives are manifold learning and probability distribution transformation \cite{63}. From the OT perspective, the generator in GAN is responsible for computing a transport map, while the discriminator measures the Wasserstein distance between the generated distribution and the real distribution. An \emph{et al.} \cite{63} proposed an AE-OT-GAN model that overcame mode collapse and incorporated explicit correspondence between latent codes and real images by applying a semi-discrete OT. In this paper, we combine OT and deep learning methods for improved interpretability and enhanced performance in generative models.

\begin{figure*}[!htbp]
\centerline{\includegraphics[width=1.7\columnwidth]{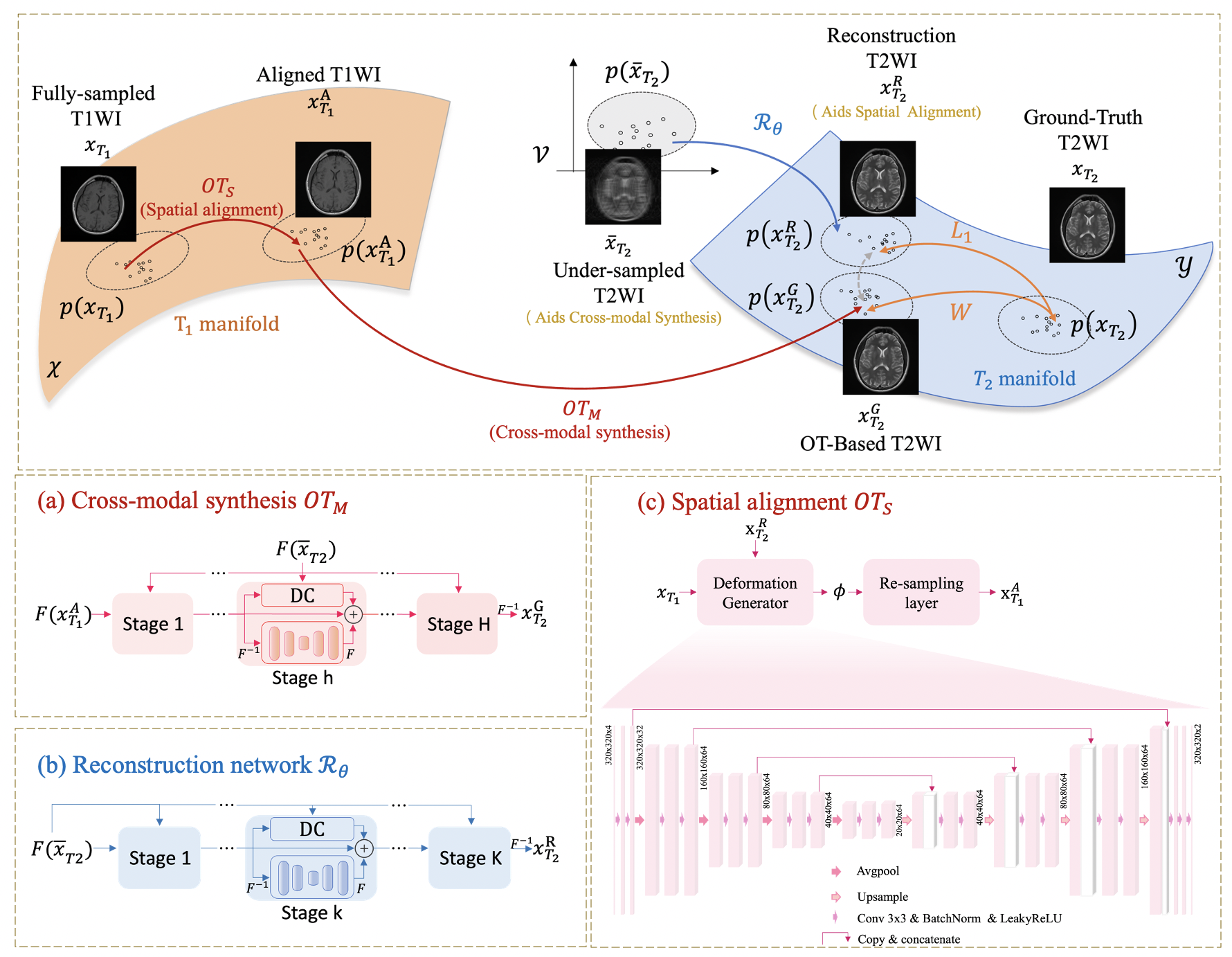}}
\caption{The exhibition of the designed framework. The reconstruction network $R_{\theta}$ uses the under-sampled T2WI $\overline{x}_{T2}$ to obtain the reconstructed T2WI $p(x_{T2}^{R})$ concentrated on $T2$ manifold. The OT process about the spatial alignment module $OT_{\mathcal{S}}$ maps the distribution of the full-sampled T1WI $p(x_{T1})$ to the distribution of the aligned T1WI $p(x_{T1}^A)$ in $T1$ manifold, which is closer to the distribution of the reconstructed T2WI $p(x_{T2}^{R})$. The OT process about the cross-modal synthesis module $OT_{\mathcal{M}}$ transports the distribution of the aligned T1WI $p(x_{T1}^A)$ to the distribution of the synthetic T2WI $p(x_{T2}^G)$ in $T2$ manifold, which is similar to the distribution of the ground truth T2WI $p(x_{T2})$.}
\label{fig2}
\end{figure*}

\section{Methods}

In Section \ref{OT}, two OT processes regarding the spatial alignment module and the cross-modal synthesis module are represented. Secondly, Section \ref{re} demonstrates that the proposed framework ensures the reconstructed T2WIs are close to the synthetic T2WIs on the T2 image manifold. Finally, we display the training scheme in Section \ref{tr}.

\subsection{T1WI-aided Cross-modal Synthesis Process}
\label{OT}

T1WIs and T2WIs provide complementary information about the same interest region of human beings. Inspired by the same underlying information and latent feature correlation among them, the target image reconstruction benefits from fully-sampled source images. In this paper, the fully-sampled T1WIs $x_{T1}$ are used to synthesize the fully-sampled T2WIs $x_{T2}^G$.

The proposed T1WI-based cross-modal synthesis process is expressed as:
\begin{equation}
\label{eq5}
x_{T2}^{G}=\mathcal{G}_{\varphi}(x_{T1}),
\end{equation}
where $\mathcal{G}_{\varphi}$ is the cross-modal synthesis network with the parameters $\varphi$.

However, as shown in Fig. \ref{fig1}, the spatial misalignment between paired images from different modalities is inevitable, which brings the negative transfer of the auxiliary information and reduces the reconstruction quality. To alleviate the negative transfer, before the cross-modal synthesis process, we design a spatial alignment module to align the auxiliary T1WIs with the under-sampled target T2WIs.

Finally, the T1WI-based cross-modal synthesis mapping $\mathcal{G}_{\varphi}$ is decomposed into two OT processes, the spatial alignment mapping  $OT_{\mathcal{S}}$ of $T1$ modality and the cross-modal synthesis mapping $OT_{\mathcal{M}}$ from the aligned T1WI $x_{T1}^A$ to the synthetic T2WI $x_{T2}^G$. Thus, the T1WI-based cross-modal synthesis mapping $\mathcal{G}_{\varphi}$, which is formulated as:
\begin{equation}
\label{eq6}
\mathcal{G}_{\varphi}(x_{T1})=OT_{\mathcal{M}} \circ OT_{\mathcal{S}}(x_{T1}).
\end{equation}

This method employs an iterative strategy for the optimization of $ \mathcal{G}_{\varphi} $. Initially, a global parameter update for the synthesis mapping $ \varphi $ is conducted. Subsequently, we undertake $ \mathcal{I} $ iterations of refinement, focusing on the sub-problems $ OT_{\mathcal{S}} $ and $ OT_{\mathcal{M}} $. Each utilizing their respective loss functions. The purpose of delineating $ \mathcal{I} $ iterations is to precisely tune the individual components within their specific parameter spaces.

During the initial phase of each iteration, we focus on optimizing $ OT_{\mathcal{M}} $,  whose network parameters are denoted by $m$.  Next, we fix the parameters $m$ and optimize $ OT_{\mathcal{S}} $, with its network parameters denoted by $s$. This staged refinement process aims to incrementally enhance the model's efficacy in cross-modal synthesis.

An OT problem can be described as that of weighing the differences between two probability distributions, which is primarily proposed by Monge in 1781 \cite{75}. The transport plan aims to minimize the cost of warping an input probability distribution onto another. In 1942, Kantorovich assumed that a source would be transported onto several targets, which promotes further development \cite{76}.

\subsubsection{An OT-view of Spatial Alignment Module}
\label{net}
This work introduces an optimal transport (OT) perspective on spatial alignment mapping. Suppose $p(x_{T1})$ is the probability measure of a T1-weighted image (T1WI). In accordance with the Manifold Distribution Hypothesis \cite{125}, the distribution of a specific class of natural data is concentrated on a low-dimensional manifold within a high-dimensional data space. Consequently, the probability measure is concentrated on a manifold, specifically the T1 manifold, embedded in the space $\mathcal{X}$. The probability distribution transformation map $OT_{\mathcal{S}}$ is designed to transport the probability measure $p(x_{T1})$ to the probability measure $p(x_{T1}^A)$ concentrated on the T1 manifold embedded in the space $\mathcal{X}$. In our optimization strategy, a total of $ \mathcal{I} $ iterations are executed. At each iteration $ i $, we define the OT problem as follows:

\begin{equation}
\label{eq7}
\begin{aligned}
    \mathcal{L}_{\mathcal{G}_{\varphi}} = & \min_{OT_{\mathcal{S}}^{(i)}} \int_{\mathcal{X}} c\Big(x_{T1},OT_{\mathcal{M}}^{(i)} \circ OT_{\mathcal{S}}^{(i)}(x_{T1})\Big) d p(x_{T1})\\
    \text{s. t. }  & OT_{\mathcal{M}}^{(i)} \circ OT_{\mathcal{S}}^{(i)}\#p(x_{T1}) = p(x_{T2}).
\end{aligned}
\end{equation}

where $ OT_{\mathcal{M}}^{(i)} $ represents the parameters of the cross-modal synthesis mapping that have been individually optimized for $ i $ iterations and are now fixed. $OT_{\mathcal{S}}^{(i)}\#p(x_{T1}) = p(x_{T1}^{A^{(i)}}) $ ensures that the distribution of the spatially aligned T1-weighted images  $ x_{T1}^{A^{(i)}}  $. If a minimum $OT_{\mathcal{S}}^*$ exists, it is named as the OT map. According to convex optimization theory, Eq. (\ref{eq7}) can be relaxed to the Kantorovich problem. Its dual form is as follows \cite{15}:

\begin{equation}
\label{eq17}
\begin{aligned}
\min_{OT_\mathcal{S}^{(i)}}\max_{\psi_\gamma} & \int\psi_\gamma(OT_{\mathcal{M}}^{(i)} \circ OT_{\mathcal{S}}^{(i)}(x_{T1}))dp(x_{T1}) \\ 
+ & \int \psi^c_\gamma(x_{T2})dp(x_{T2}) 
\end{aligned}
\end{equation}

In the spatial alignment task, the discriminator $\mathcal{D}_{\gamma}$, denoted by $\psi_\gamma$, enhances the estimation of the Wasserstein distance. This distance is between the synthetic T2-weighted image $ x_{T2}^G$ and the ground truth image $ x_{T2} $. The aim is to indirectly measure how well the images align . A smaller distance means a better alignment. The discriminator $\psi_\gamma $ also serves as the Kantorovich potential. $\psi_\gamma $ satisfies the 1-Lipschitz condition, which is typically ensured via spectral normalization \cite{124}. For $ L^1 $ transportation cost, its c-transform exhibits the relationship $ \psi_\gamma^c = -\psi_\gamma $. Through an iterative optimization algorithm between maximization and minimization, we can numerically obtain both the map $ OT_\mathcal{S}^{(i)} $ and the maximizer $ \psi_\gamma^* $ \cite{82}.

\subsubsection{An OT-view of the Cross-modal Synthesis Module}

This work also presents an OT view of cross-modal synthesis module. $p(x_{T1}^{A^{(i)}})$ concentrated on the $T1$ manifold is the probability measure in the space $\mathcal{X}$. The goal of the probability distribution transformation map $OT_{\mathcal{M}}^{(i+1)}$ is to convert the probability measure of the aligned T1WI $p(x_{T1}^{A^{(i)}})$ into the probability measure of the ground truth T2WI $p(x_{T2})$, which is focused on the $T2$ manifold embedded in the space $\mathcal{Y}$.

Suppose a map $OT_{\mathcal{M}}^{(i+1)}: \mathcal{X} \rightarrow \mathcal{Y}$ transports the probability measure $p(x_{T1}^{A^{(i)}})$ into the probability measure $p(x_{T2})$. The OT problem is formulated as:

\begin{equation}
\label{eq8}
\begin{aligned}
    \mathcal{L}_{OT_{\mathcal{M}}} &= \min_{OT_{\mathcal{M}}^{(i+1)}} \int_{\mathcal{X}} c\Big(x_{T1}^{A^{(i)}}, OT_{\mathcal{M}}^{(i+1)}(x_{T1}^{A^{(i)}})\Big) d p(x_{T1}^{A^{(i)}}) \\
    \text{s. t. } & OT_{\mathcal{M}}^{(i+1)} \#p(x_{T1}^{A^{(i)}}) = p(x_{T2}).
\end{aligned}
\end{equation}

Similarly, the Kantorovich dual form is as follows:
\begin{equation}%
\label{eq18}
\begin{aligned}
\min_{OT_\mathcal{M}^{(i+1)}}\max_{\psi_\omega} & \int\psi_\omega(OT_\mathcal{M}^{(i+1)}(x_{T1}^{A^{(i)}}))dp(x_{T1}^{A^{(i)}}) \\
+ & \int\psi^c_\omega(x_{T2})dp(x_{T2})
\end{aligned}
\end{equation}

In the task of cross-modal synthesis, the discriminator $\mathcal{D}_{\omega}$ refines the approximation of the Wasserstein distance $ \psi_\omega $. This distance is measured between the synthetic T2WI $ x_{T2}^G $ and the ground truth T2WI $ x_{T2} $. The synthetic image $ x_{T2}^G $ is produced by the cross-modal synthesis module, which is achieved by applying map $ OT_{\mathcal{M}}^{(i+1)} $ to the aligned T1WI $ x_{T1}^{A^{(i)}} $. Subsequently, the same optimization method is employed to numerically obtain the map $ OT_{\mathcal{M}}^{(i+1)} $ and the maximizer $ \psi_\omega^* $.

\subsection{The Complementarity between Two Tasks}
\label{re}

When the high-quality reconstructed T2WI $x_{T2}^R$ is achieved, the aligned T1WI $x_{T1}^A$ is more close to the ground truth $x_{T2}$. At the same time, the synthetic T2WI $x_{T2}^G$ has shorter distance to the ground truth $x_{T2}$ in $T2$ manifold. In the proposed framework, we also aim to making the distance between the distribution of the synthetic T2WI $x_{T2}^G$ and the reconstructed T2WI $x_{T2}^R$ short by minimizing their distance to the ground truth T2WI $x_{T2}$. Therefore, the reconstruction task and the cross-modal task benefit each other, which is presented in the following theorem.

\newtheorem{theorem}{Theorem}

\begin{theorem}
Let $x_{T2}^{R}$, $x_{T2}^{G}$ be the reconstructed T2WI and the synthetic T2WI, and $x_{T2}$ be the full sampled T2WI, then we have the following inequality held:
\begin{equation}\label{thm}
\|x_{T2}^{R}-x_{T2}^{G}\|_{L_1} \leq \|x_{T2}^{R}-x_{T2}\|_{L_1} + C \|x_{T2}^{G}-x_{T2}\|_{W_1},
\end{equation}
where $\|\cdot\|_{L_1}$ and $\|\cdot\|_{W_1}$ are $L_1$ and $1$-Wasserstein norms, respectively, and the $C>0$ is a constant.
\end{theorem}

\noindent{Proof:} Let $\Omega \subset R^3$ be the region of images, and it is compact. It is a natural assumption that $x_{T2}, x_{T2}^{R}, x_{T2}^{G}, x_{T1}, x_{T1}^{A} \in BV(\Omega)$, where $BV(\Omega)$ is the set of all bounded variation functions on $\Omega$. Then, it is easily validated that $(BV(\Omega), L_1)$, is a metric space with $L_1$-metric. Therefore, from the triangle inequality of the distance, we have

\[
\|x_{T2}^{R}-x_{T2}^{G}\|_{L_1} \leq \|x_{T2}^{R}-x_{T2}\|_{L_1}+ \|x_{T2}^{G}-x_{T2}\|_{L_1}.
\]

On the other hand, because $\Omega$ is compact, from Remark 8.2 \cite{15}, there is $\|x_{T2}^{G}-x_{T2}\|_{W_1} \leq diam(\Omega) \|x_{T2}^{G}-x_{T2}\|_{L_1}$, where diam($\Omega$) is the diameter of the set $\Omega$. Therefore, we have the inequality (\ref{thm}) held, where the constant $C = \frac{1}{diam(\Omega)}$.

\noindent{\bf{Remark:}} It is remarkable that the inequality (\ref{thm}) gives an explanation of complementarity between two tasks. That is, the error between the reconstructed T2WIs and the generated T2WIs is controlled by the errors of two tasks. In other words, with the iteration increasing, $x_{T2}^{R}$ and $x_{T2}^{G}$ become more and more close. Due to the alternative updating, the better reconstructed result improves the generating process, while the better generated result enhances the reconstructing process. It actually realizes two tasks complement each other.


\subsection{The Training Scheme of the Proposed Method}
\label{tr}

As shown in Fig. \ref{fig4}, the reconstruction task under the deep learning framework in this paper is formulated as:

\begin{equation}
\label{eq4}
\mathcal{L}_{Rec} = \min_{\theta} \|x_{T2}- \mathcal{R}_{\theta}(\overline{x}_{T2})\|_1,
\end{equation}
where $\mathcal{R}_{\theta}$ is a T2WI reconstruction network which restores the unclear target images. $\theta$ are the optimized parameters of the reconstruction network.

Similar to the work \cite{8}, in the spatial alignment process $OT_{\mathcal{S}}$, if we blindly pursue the maximum similarity between the two distributions $p(x_{T1}^A)$ and $p(x_{T2}^R)$, it will lead to image distortion. Thus, a regularization term is introduced to avoid extremely deforming the image, that is:

\begin{equation}
\label{eq9}
\mathcal{L}_{Reg} = \sum_{a\in S(b)}H(a,b)\|\phi(a)-\phi(b)\|,
\end{equation}
where $S(b)$ represents neighboring pixels of a pixel $b$. $H(a,b) = e^{-\|{x_{T1}^A[a]-x_{T1}^A[b]\|}}$ is a bilateral filter which is introduced to prevent over-smoothness. $\phi$ is the estimated displacement field. Finally, in this paper the overall function of loss is formulated as

\begin{equation}
\label{eq10}
\begin{aligned}
    \min_{\theta, \varphi, \gamma, \omega} \alpha \mathcal{L}_{Rec} + \beta \mathcal{L}_{
    \mathcal{G}_{\varphi}} + \delta \mathcal{L}_{OT_{\mathcal{M}}} + \eta \mathcal{L}_{Reg},
\end{aligned}
\end{equation}
where $\alpha, \beta, \delta, \eta$ are hyperparameters that need to be adjusted.

The detailed training procedure of the cross-modal reconstruction is outlined in Algorithm 1.

\renewcommand{\algorithmicrequire}{\textbf{Input:}}
\renewcommand{\algorithmicensure}{\textbf{Output:}}

\begin{algorithm}
    \caption{The Proposed Cross-modal Reconstruction Method}
    \begin{algorithmic}[1]
        \REQUIRE Paired fully-sampled T1WIs $x_{T1}$ and fully-sampled T2WIs $x_{T2}$.
        \ENSURE Optimized neural networks, $\mathcal{R}_{\theta^*}, \mathcal{G}_{\varphi^*}, \mathcal{D}_{\gamma^*}, \mathcal{D}_{\omega^*}$.
        \REPEAT
        \STATE $x_{T2}^{R} \leftarrow \mathcal{R}_{\theta}\left(\bar{x}_{T2}\right)$;
        \STATE $x_{T1}^{A} \leftarrow OT_{\mathcal{S}}\left(x_{T1}, x_{T2}^{R} \right)$;
        \STATE $x_{T2}^{G} \leftarrow OT_{\mathcal{M}}\left(x_{T1}^{A} \right)$;
        \STATE Use Eqs. (\ref{eq17}), (\ref{eq18}), (\ref{eq4}) and (\ref{eq9}) to calculate $\mathcal{L}_{
    \mathcal{G}_{\varphi}}$, $\mathcal{L}_{OT_{\mathcal{M}}}$, $\mathcal{L}_{Rec}$ and $\mathcal{L}_{Reg}$, respectively.
        \STATE $\theta \leftarrow \theta-\epsilon \partial_{\theta}\left(\alpha \mathcal{L}_{Rec} + \beta \mathcal{L}_{
    \mathcal{G}_{\varphi}} + \delta \mathcal{L}_{OT_{\mathcal{M}}} + \eta \mathcal{L}_{Reg}\right)$
        \STATE $\varphi \leftarrow \varphi-\epsilon \partial_{\varphi}\left(\alpha \mathcal{L}_{Rec} + \beta \mathcal{L}_{
    \mathcal{G}_{\varphi}} + \delta \mathcal{L}_{OT_{\mathcal{M}}} + \eta \mathcal{L}_{Reg}\right)$
        \FOR{$i = 1$ to $\mathcal{I}$} 
            \STATE $m \leftarrow m - \epsilon \partial_{m} \mathcal{L}_{OT_{\mathcal{M}}}$
            \STATE $\omega \leftarrow \omega-\epsilon \partial_{\omega}\left(\mathcal{D}_{\omega}(x_{T2}) - \mathcal{D}_{\omega}(x_{T2}^{G^{(i-1)}})\right)$
            \STATE $x_{T2}^{G^{(\frac{2i-1}{2})}} \leftarrow OT_{\mathcal{M}}^{(i)} \left(x_{T1}^{A^{(i-1)}} \right)$
            \STATE $s \leftarrow s - \epsilon \partial_{s} \mathcal{L}_{
    \mathcal{G}_{\varphi}}$
           \STATE $\gamma \leftarrow \gamma-\epsilon \partial_{\gamma}\left(\mathcal{D}_{\gamma}(x_{T2}) - \mathcal{D}_{\gamma}(x_{T2}^{G^{(\frac{2i-1}{2})}}) \right)$
            \STATE $x_{T1}^{A^{(i)}} \leftarrow OT_{\mathcal{S}}^{(i)} \left(x_{T1} \right)$
            \STATE $x_{T2}^{G^{(i)}} \leftarrow OT_{\mathcal{M}}^{(i)} \left(x_{T1}^{A^{(i)}} \right)$
        \ENDFOR
        \UNTIL Convergence
    \end{algorithmic}
\end{algorithm}

\section{Experiments}
\label{sec:experiments}
In this Section, we first introduce the datasets in Section \ref{Details}. Implementation details and baselines are presented in Section \ref{Base}. In Section \ref{Results}, we summarize and analyze the experimental results from the view of quantitative and qualitative evaluation. Finally, Section \ref{discussion} provides an ablation study to investigate the effectiveness of the proposed cross-modal strategies.

\subsection{Experimental Datasets}
\label{Details}

The proposed framework aims to reconstruct T2-weighted axial brain MRIs using fully-sampled T1-weighted MRIs, taking advantage of the complementary information provided by these modalities. In this section, we evaluate the performance of the proposed method in two MRI datasets: FastMRI, which consists of partial brain MRI scans, and an in-house dataset comprising whole brain MRI scans. Both datasets are acquired from Siemens scanners and are available in Digital Imaging and Communications in Medicine (DICOM) and Neuroimaging Informatics Technology Initiative (NIfTI) formats, respectively. To handle the 3D MRI volumes, we decompose them into 2D slices and treat them as inputs to the framework. In the FastMRI dataset, we adopt the same selection criteria as the reference \cite{8}, which includes 340 pairs of T1 and T2 weighted axial brain MRIs. Specifically, we use 170 pairs of volumes (2720 pairs of slices) for training, 68 pairs of volumes (1088 pairs of slices) for validation, and 102 pairs of volumes (1632 pairs of slices) for testing. The in-plane size of T1WIs and T2WIs is 320 $\times$ 320, with a resolution of 0.68 $\times$ 0.68 and a slice spacing of $5mm$.

The in-house dataset consists of 36 subjects' whole brain MRI scans, including 3D paired T1WIs (TE=$1.4e+02ms$; Time=$161149.853ms$) and T2WIs (TE=$1.2e+02ms$; Time=$163011.360ms$). Out of these, 30 pairs of volumes (2097 pairs of slices) are used for training, 3 pairs of volumes (225 pairs of slices) for validation, and 3 pairs of volumes (225 pairs of slices) for testing. The in-plane spacing is 0.68mm $\times$ 0.68mm, and the slice thickness is 3mm. Each slice has a spatial size of 320 $\times$ 320.

\subsection{Baselines and Implementation Details}
\label{Base}

The undersampled magnetic resonance images are obtained by applying a sampling ratio of $25\%$, $12.5\%$, or $6.25\%$ to mask the $k$-space. We consider three sampling patterns: random, equispaced, and radial. For random and equispaced sampling, $32\%$ of low-spatial-frequency data captures the general shape of organs. The rest focuses on finer details with high-spatial-frequency data. The radial sampling method is different. It captures data along radial lines in the $k$-space. This method is effective for capturing structural details at various sampling densities.

The reconstruction network, denoted as $R_{\theta}$, is constructed using cascades of E2E-VarNet architectures \cite{66}. Each E2E-VarNet architecture comprises four down-sampling CNN layers in the encoder and four up-sampling CNN layers in the decoder. The spatial alignment module consists of a deformation generator and a re-sampling layer, with the deformation field generator utilizing four down-sampling CNN layers in the encoder and four up-sampling CNN layers in the decoder. The cross-modal synthesis process, denoted as $OT_{\mathcal{M}}$, employs the same architecture as the reconstruction network. The discriminator networks $\mathcal{D}{\gamma}$ and $\mathcal{D}{\omega}$ take complex-valued inputs represented by two channels and output a real-valued scalar indicating the similarity between the magnetic resonance images. The Adam optimizer with a learning rate of $1 \times 10^{-4}$ is used for all networks. The reconstruction loss $\mathcal{L}_{Rec}$, spatial alignment loss $\mathcal{L}_{\mathcal{G}_{\varphi}}$, cross-modal image synthesis loss $\mathcal{L}_{OT_{\mathcal{M}}}$, and smoothness loss $\mathcal{L}_{Reg}$ are weighted by $3000$, $1$, $1$, and $1000$, respectively. The experiments are trained until convergence on a NVIDIA GeForce RTX 2080 Ti.

For the comparison, we evaluate the proposed OT-based model against several competing methods: SAN\cite{8}, MTrans\cite{4}, MT-NET\cite{113}, and MC-CDic\cite{114}. SAN addresses spatial discrepancies in cross-modal reconstruction. MTrans uses an improved multi-head attention mechanism and a cross attention module for rich multi-modal information capture. MT-NET employs an Edge-preserving masked autoencoder for contextual and structural information from a reference modality in cross-modal image synthesis. MC-CDic combines convolutional dictionary learning with deep unfolding techniques for multi-contrast MRI super-resolution and reconstruction. It shows effectiveness in diverse MRI contrasts and image resolution enhancement. In our approach, we aim to combine spatial alignment with cross-modal synthesis within the OT framework. This effort seeks to enhance alignment accuracy and potentially bridge modal differences.

To assess the reconstruction performance accurately and comprehensively, we utilize Peak Signal-to-Noise Ratio (PSNR) \cite{84}, Structural Similarity Index Metric (SSIM) \cite{79}, and Normalized Mean Square Error (NMSE) \cite{83} as metrics to measure the discrepancy between the reconstructed T2WI and the real one.

\subsection{Experiments on Two MRI Datasets}
\label{Results}
\subsubsection{Quantitative Evaluation}

\begin{table}[!htbp]
\renewcommand\arraystretch{1}
\caption{Average (with standard deviation) reconstruction results, in terms of PSNR, SSIM and NMSE on FastMRI with different masks. The best result is in bold.}
    \begin{center}
        \setlength{\tabcolsep}{1.4pt}
        \renewcommand\arraystretch{1.4}
        \footnotesize
        \begin{tabular}{cccccc}
            \toprule
            \rowcolor{gray!20}
            \multicolumn{1}{c}{\textbf{Ratio}} & \textbf{Masks} & \textbf{Methods} & \textbf{PSNR} & \textbf{SSIM} & \textbf{NMSE} \\
            \midrule
            \multirow{15}*{25\%}
            & \multirow{5}*{\rotatebox{90}{Random}}
            & Zero-filling & 26.97$\pm$1.16 & 0.7282$\pm$0.0360 & 0.0492$\pm$0.0286 \\
            & & MTrans \cite{4} & 35.34$\pm$1.39 & 0.9448$\pm$0.0071 & 0.0097$\pm$0.0013\\
            & & MC-CDic \cite{114} & 39.92$\pm$2.85 & 0.9702$\pm$0.0156 & 0.0047$\pm$0.0084 \\
            & & SAN \cite{8} & 43.91$\pm$1.94 & \textbf{0.9888}$\pm$\textbf{0.0053} & 0.0019$\pm$0.0007 \\
            & &OT-Based & \textbf{44.23}$\pm$\textbf{1.32} & 0.9882$\pm$0.0031 & \textbf{0.0012}$\pm$\textbf{0.0005} \\
            \cmidrule{2-6}
            & \multirow{5}*{\rotatebox{90}{Equispaced}}
            & Zero-filling & 26.79$\pm$1.17 & 0.7321$\pm$0.0377 & 0.0515$\pm$0.0302 \\
            & & MTrans \cite{4} & 34.28$\pm$1.35 & 0.9593$\pm$0.0080 & 0.0121$\pm$0.0303\\
            & & MC-CDic \cite{114} & 40.78$\pm$2.97 & 0.9771$\pm$0.0130 & 0.0051$\pm$0.0091 \\
            & & SAN \cite{8} & 40.81$\pm$1.92 & \textbf{0.9821}$\pm$\textbf{0.0069} & 0.0040$\pm$0.0009 \\
            & &OT-Based & \textbf{41.46}$\pm$\textbf{0.44} & 0.9809$\pm$0.0007 & \textbf{0.0024}$\pm$\textbf{0.0005} \\
            \cmidrule{2-6}
            & \multirow{5}*{\rotatebox{90}{Radial}}
            & Zero-filling & 33.37$\pm$1.44 & 0.6952$\pm$0.1404 & 0.0218$\pm$0.0140 \\
            & & MTrans \cite{4} & 36.82$\pm$1.28 & 0.9623$\pm$0.0210 & 0.0081$\pm$0.0043\\
            & & MC-CDic \cite{114} & 43.01$\pm$1.00 & 0.9812$\pm$0.0107 &0.0026$\pm$0.0077 \\
            & & SAN \cite{8} & 45.55$\pm$1.42 & 0.9902$\pm$0.0047 & \textbf{0.0020}$\pm$\textbf{0.0161} \\        
            & &OT-Based & \textbf{45.66}$\pm$\textbf{2.44} & \textbf{0.9905}$\pm$\textbf{0.0045} & 0.0024$\pm$0.0005 \\
            \midrule
            \multirow{15}{*}{12.5\%}
            & \multirow{5}{*}{\rotatebox{90}{Random}}
            & Zero-filling & 23.60$\pm$1.82 & 0.5462$\pm$0.0490 & 0.0991$\pm$0.0474 \\
            & & MTrans \cite{4} & 33.18$\pm$1.34 & 0.9318$\pm$0.0201 & 0.0154$\pm$0.0037\\
            & & MC-CDic \cite{114} & 35.79$\pm$2.92 & 0.9498$\pm$0.0267 & 0.0117$\pm$0.0122 \\
            & & SAN \cite{8} & 38.06$\pm$2.04 & 0.9729$\pm$0.0103 & 0.0056$\pm$0.0007 \\
            & &OT-Based & \textbf{39.38}$\pm$\textbf{1.00} & \textbf{0.9744}$\pm$\textbf{0.0037} & \textbf{0.0031}$\pm$\textbf{0.0003} \\
            \cmidrule{2-6}
            & \multirow{5}{*}{\rotatebox{90}{Equispaced}}
            & Zero-filling & 24.76$\pm$1.81 & 0.5428$\pm$0.0488 & 0.0956$\pm$0.0458 \\
            & & MTrans \cite{4} & 32.10$\pm$1.25 & 0.9240$\pm$0.0172 & 0.0178$\pm$0.0122\\
            & & MC-CDic \cite{114} & 36.64$\pm$2.19 & 0.9561$\pm$0.0242 & 0.0099$\pm$0.0142 \\
            & & SAN \cite{8} & 39.38$\pm$2.06 & \textbf{0.9775}$\pm$\textbf{0.0088} & 0.0036$\pm$0.0010 \\
            & &OT-Based & \textbf{39.94}$\pm$\textbf{1.12} & 0.9765$\pm$0.0026 & \textbf{0.0035}$\pm$\textbf{0.0006}\\
            \cmidrule{2-6}
            & \multirow{5}*{\rotatebox{90}{Radial}}
            & Zero-filling & 29.44$\pm$1.88 & 0.5675$\pm$0.0649 & 0.0533$\pm$0.0286 \\
            & & MTrans \cite{4} & 36.20$\pm$1.90 & 0.9427$\pm$0.0209 & 0.0092$\pm$0.0402\\
            & & MC-CDic \cite{114} & 39.75$\pm$2.14 & 0.9694$\pm$0.0180 & 0.0051$\pm$0.0107 \\
            & & SAN \cite{8} & 41.85$\pm$1.37 & 0.9824$\pm$0.0083 & 0.0036$\pm$0.0130 \\
            & &OT-Based & \textbf{42.52}$\pm$\textbf{2.37} & \textbf{0.9838}$\pm$\textbf{0.0078} & \textbf{0.0032}$\pm$\textbf{0.0117} \\
            \midrule
            \multirow{15}{*}{6.25\%}
            & \multirow{5}{*}{\rotatebox{90}{Random}}
            & Zero-filling & 24.43$\pm$1.87 & 0.4850$\pm$0.0552 & 0.1500$\pm$0.0137 \\
            & & MTrans \cite{4} & 30.78$\pm$1.78 & 0.9193$\pm$0.0106 & 0.0206$\pm$0.0078\\
            & & MC-CDic \cite{114} & 34.38$\pm$2.07 & 0.9434$\pm$0.0287 & 0.0162$\pm$0.0068 \\
            & & SAN \cite{8} & 33.91$\pm$1.28 & 0.9488$\pm$0.0073 & 0.0121$\pm$0.0011 \\
            & &OT-Based & \textbf{36.30}$\pm$\textbf{1.27} & \textbf{0.9512}$\pm$\textbf{0.0061} & \textbf{0.0089}$\pm$\textbf{0.0004} \\
            \cmidrule{2-6}
            & \multirow{5}{*}{\rotatebox{90}{Equispaced}}
            & Zero-filling & 23.96$\pm$1.56 & 0.4589$\pm$0.1491 & 0.1639$\pm$0.0529 \\
            & & MTrans \cite{4} & 29.63$\pm$1.35 & 0.8922$\pm$0.0103 & 0.0273$\pm$0.0069 \\
            & & MC-CDic \cite{114} & 34.78$\pm$2.69 & 0.9466$\pm$0.0274 & 0.0151$\pm$0.0172 \\
            & & SAN \cite{8} & 34.76$\pm$1.74 & 0.9474$\pm$0.0062 &0.0175$\pm$0.0006 \\
            & & OT-Based & \textbf{35.45}$\pm$\textbf{1.42} & \textbf{0.9573}$\pm$\textbf{0.0097} & \textbf{0.0143}$\pm$\textbf{0.0087} \\
            \cmidrule{2-6}
            & \multirow{5}*{\rotatebox{90}{Radial}}
            & Zero-filling & 26.79$\pm$1.68 & 0.4773$\pm$0.1632 & 0.0937$\pm$0.0438 \\
            & & MTrans \cite{4} & 32.13$\pm$1.56 & 0.9103$\pm$0.0520 & 0.0103$\pm$0.0398\\
            & & MC-CDic \cite{114} & 37.41$\pm$2.93 & 0.9596$\pm$0.0212 & 0.0089$\pm$0.0198 \\
            & & SAN \cite{8} & 38.60$\pm$2.36 & 0.9703$\pm$0.0136 & 0.0069$\pm$0.0188 \\
            & &OT-Based & \textbf{39.92}$\pm$\textbf{2.31} & \textbf{0.9749}$\pm$\textbf{0.0123} & \textbf{0.0053}$\pm$\textbf{0.0145} \\
            \bottomrule
        \end{tabular}
        \label{tb1}
    \end{center}
\end{table}

\begin{figure}[!htbp]
\centerline{\includegraphics[width=\columnwidth]{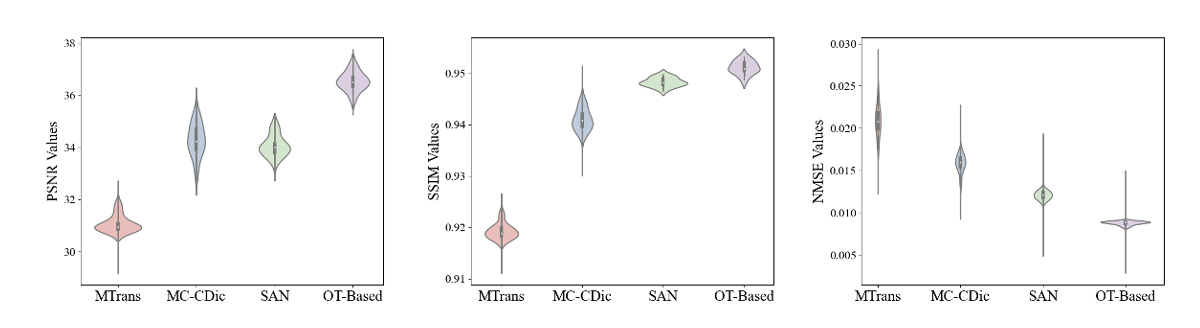}}
\caption{The provided violin plot visually depicts the results of the Fast dataset obtained through random sampling at a ratio of {6.25\%} using random masks. The plot showcases three distinct violin-shaped distributions, each representing the outcomes for PSNR, SSIM, and NMSE obtained through different methods.}
\label{fig3}
\end{figure}

We evaluate the performance of the proposed cross-modal reconstruction algorithms using metrics such as PSNR, SSIM, and NMSE, which quantify the similarity between the restored image and the fully-sampled ground truth image. The results of the reconstruction on two magnetic resonance image datasets under various under-sampling patterns are presented in Tables \ref{tb1} and \ref{tb2}. In these tables, the best result is highlighted in bold.

Our analysis reveals that the OT-based method consistently demonstrates superior performance at sample ratios of $12.5\%$ and $25\%$. Even at a significantly low sample ratio of $6.25\%$, the OT-based method outperforms other techniques, exhibiting substantially higher PSNR and SSIM values, as well as lower NMSE values.

While the SSIM of the OT-based method may be slightly lower than that of SAN \cite{8} and MC-CDic \cite{114}, it is important to note that the OT-based method exhibits a smaller standard deviation. Consequently, the improvement in PSNR is more effective. For instance, on the FastMRI dataset with a sample ratio of $12.5\%$ using a random pattern, the OT-based method enhances the PSNR result from 38.06 dB to 39.38 dB. From Fig. \ref{fig3}, we can see that even if the sampling rate is very low, we can still obtain good reconstruction results.

\begin{table}[!htbp]
\renewcommand\arraystretch{1}
\caption{Average (with standard deviation) reconstruction results, in terms of PSNR, SSIM and NMSE on in-house with different masks. The best result is in bold.}
    \begin{center}
        \setlength{\tabcolsep}{1.2pt}
        \renewcommand\arraystretch{1.4}
        \footnotesize
        \begin{tabular}{cccccc}
            \toprule
            \rowcolor{gray!20}
            \multicolumn{1}{c}{\textbf{Ratio}} & \textbf{Masks} & \textbf{Methods} & \textbf{PSNR} & \textbf{SSIM} & \textbf{NMSE} \\
            \midrule
            \multirow{15}*{25\%} 
            & \multirow{5}*{\rotatebox{90}{Random}}
            & Zero-filling & 25.33$\pm$1.28 & 0.7738$\pm$0.0564 & 0.2386$\pm$0.0973 \\
            & &MTrans \cite{4} & 36.93$\pm$1.67 & 0.9730$\pm$0.0201 & 0.0395$\pm$0.0073 \\
            & & MC-CDic \cite{114} & 37.10$\pm$3.04 & 0.9863$\pm$0.0082 & 0.0361$\pm$0.0225 \\
            & & SAN \cite{8} & 38.12$\pm$1.81 & 0.9940$\pm$0.0072 & 0.0061$\pm$0.0007 \\
            & & OT-Based & \textbf{38.48$\pm$1.76} & \textbf{0.9946$\pm$0.0023} & \textbf{0.0057$\pm$0.0008} \\
            \cmidrule{2-6}
            & \multirow{5}*{\rotatebox{90}{Equispaced}}
            & Zero-filling & 25.27$\pm$1.33 & 0.7799$\pm$0.0545 & 0.2433$\pm$0.1022 \\
            & & MTrans \cite{4} & 32.50$\pm$2.10 & 0.9763$\pm$0.0135 & 0.0609$\pm$0.0097\\
            & & MC-CDic \cite{114} & 35.02$\pm$2.26 & 0.9882$\pm$0.0068 & 0.0332$\pm$0.0235 \\
            & & SAN \cite{8} & 34.99$\pm$1.79 & 0.9898$\pm$0.0041 & 0.0416$\pm$0.0073 \\
            & & OT-Based & \textbf{35.87$\pm$1.51} & \textbf{0.9922$\pm$0.0032} & \textbf{0.0306$\pm$0.0015} \\
            \cmidrule{2-6}
            & \multirow{5}*{\rotatebox{90}{Radial}}
            & Zero-filling & 25.94$\pm$1.68 & 0.7143$\pm$0.0732 & 0.2178$\pm$0.1069 \\
            & & MTrans \cite{4} & 35.71$\pm$1.28 & 0.9320$\pm$0.0123 & 0.0389$\pm$0.0249\\
            & & MC-CDic \cite{114} & 37.15$\pm$2.63 & 0.9899$\pm$0.0062 & 0.0203$\pm$0.0165 \\
            & & SAN \cite{8} & 41.67$\pm$1.29 & 0.9961$\pm$0.0013 & 0.0062$\pm$0.0043 \\
            & &OT-Based & \textbf{43.81}$\pm$\textbf{1.52} & \textbf{0.9971}$\pm$\textbf{0.0011} & \textbf{0.0039}$\pm$\textbf{0.0031} \\
            \midrule
            \multirow{15}*{12.5\%} 
            & \multirow{5}*{\rotatebox{90}{Random}}
            & Zero-filling & 24.05$\pm$1.27 & 0.6552$\pm$0.0525 & 0.2993$\pm$0.0936 \\
            & & MTrans \cite{4} & 29.09$\pm$1.07 & 0.9629$\pm$0.0056 & 0.0705$\pm$0.0036 \\
            & & MC-CDic \cite{114} & 30.61$\pm$2.14 & 0.9750$\pm$0.0108 & 0.0732$\pm$0.0387 \\
            & & SAN \cite{8} & 30.70$\pm$1.34 & 0.9748$\pm$0.0029 & 0.0431$\pm$0.0093 \\
            & & OT-Based & \textbf{31.39$\pm$1.09} & \textbf{0.9825$\pm$0.0037} & \textbf{0.0348$\pm$0.0037} \\
            \cmidrule{2-6}
            & \multirow{5}*{\rotatebox{90}{Equispaced}}
            & Zero-filling & 23.91$\pm$3.14 & 0.6515$\pm$0.0510 & 0.3070$\pm$0.0933\\
            & & MTrans \cite{4} & 28.39$\pm$1.73 & 0.9566$\pm$0.0100 & 0.0806$\pm$0.0079 \\
            & & MC-CDic \cite{114} & 31.76$\pm$2.37 & 0.9796$\pm$0.0098 & 0.0578$\pm$0.0297 \\
            & & SAN \cite{8} & 31.63$\pm$1.24 & 0.9812$\pm$0.0107 & 0.0378$\pm$0.0009 \\
            & & OT-Based & \textbf{32.56$\pm$1.34} & \textbf{0.9849$\pm$0.0071} & \textbf{0.0240$\pm$0.0008} \\
            \cmidrule{2-6}
            & \multirow{5}*{\rotatebox{90}{Radial}}
            & Zero-filling & 24.54$\pm$1.85 & 0.5835$\pm$0.0231 & 0.2797$\pm$0.1142 \\
            & & MTrans \cite{4} & 31.23$\pm$2.15 & 0.9001$\pm$0.0204 & 0.0410$\pm$0.0160\\
            & & MC-CDic \cite{114} & 36.02$\pm$2.19 & 0.9882$\pm$0.0066 & 0.0240$\pm$0.0164 \\
            & & SAN \cite{8} & 37.79$\pm$1.37 & 0.9930$\pm$0.0029 & 0.0147$\pm$0.0092 \\
            & &OT-Based & \textbf{38.68}$\pm$\textbf{2.80} & \textbf{0.9939}$\pm$\textbf{0.0024} & \textbf{0.0127}$\pm$\textbf{0.0089} \\
            \midrule
            \multirow{15}*{6.25\%} 
            & \multirow{5}*{\rotatebox{90}{Random}}
            & Zero-filling & 20.70$\pm$1.26 & 0.5367$\pm$0.0497 & 0.3912$\pm$0.0757 \\
            & & MTrans \cite{4} & 23.23$\pm$2.15 & 0.9201$\pm$0.0089 & 0.2303$\pm$0.0127\\
            & & MC-CDic \cite{114} & 25.45$\pm$2.16 & 0.9441$\pm$0.0140 & 0.2159$\pm$0.0771 \\
            & & SAN \cite{8} & 26.47$\pm$1.12 & 0.9557$\pm$0.0064 & 0.1473$\pm$0.0037 \\
            & & OT-Based & \textbf{27.56}$\pm$\textbf{1.05} & \textbf{0.9653}$\pm$\textbf{0.0033} & \textbf{0.1269}$\pm$\textbf{0.0021}\\
            \cmidrule{2-6}
            & \multirow{5}*{\rotatebox{90}{Equispaced}}
            & Zero-filling & 20.59$\pm$2.23 & 0.5375$\pm$0.0022 & 0.4000$\pm$0.0741 \\
            & & MTrans \cite{4} & 26.03$\pm$1.38 & 0.9303$\pm$0.0128 & 0.2596$\pm$0.0137\\
            & & MC-CDic \cite{114} & 26.45$\pm$2.38 & 0.9645$\pm$0.0143 & 0.1743$\pm$0.0501\\
            & & SAN \cite{8} & 27.09$\pm$1.36 & 0.9617$\pm$0.0083 & 0.1694$\pm$0.0089 \\
            & & OT-Based & \textbf{27.39}$\pm$\textbf{1.27} & \textbf{0.9652}$\pm$\textbf{0.0049} & \textbf{0.1433}$\pm$\textbf{0.0014}\\
            \cmidrule{2-6}
            & \multirow{5}*{\rotatebox{90}{Radial}}
            & Zero-filling & 23.45$\pm$2.95 & 0.4699$\pm$0.0785 & 0.3394$\pm$0.1028 \\
            & & MTrans \cite{4} & 28.31$\pm$1.45 & 0.9087$\pm$0.0078 & 0.0834$\pm$0.0311\\
            & & MC-CDic \cite{114} & 33.26$\pm$2.49 & 0.9825$\pm$0.0087 & 0.0414$\pm$0.0241 \\
            & & SAN \cite{8} & 32.09$\pm$1.30 & 0.9821$\pm$0.0082 & 0.0541$\pm$0.0329 \\
            & &OT-Based & \textbf{34.16}$\pm$\textbf{2.34} & \textbf{0.9875}$\pm$\textbf{0.0057} & \textbf{0.0339}$\pm$\textbf{0.0212} \\
            \bottomrule
        \end{tabular}
        \label{tb2}
    \end{center}
\end{table}

\subsubsection{Qualitative Evaluation}
\begin{figure}[!t]
\centerline{\includegraphics[width=\columnwidth]{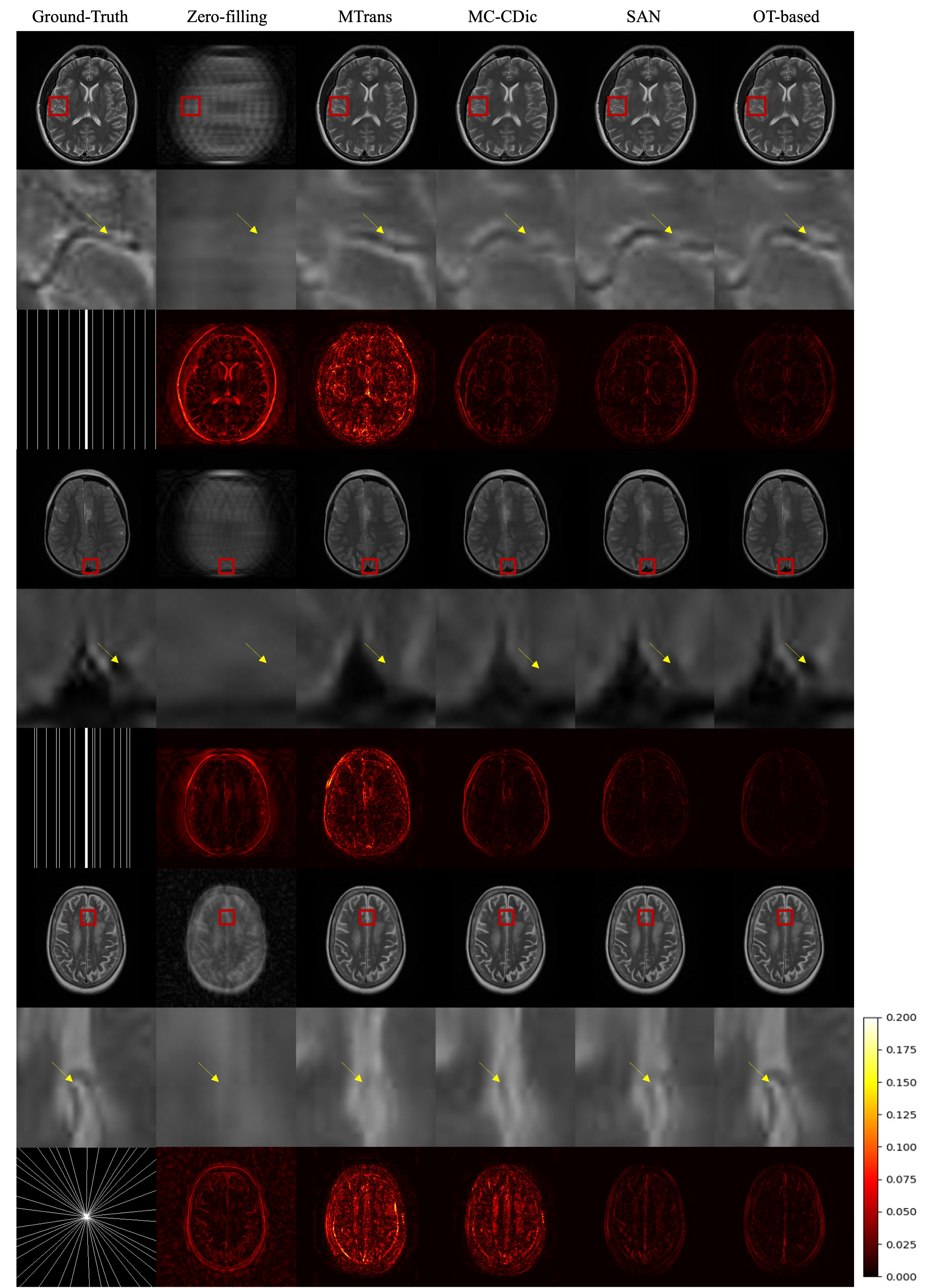}}
\caption{Comparison of the reconstruction results on the FastMRI dataset is shown. The first row shows outcomes from 16× equispaced sampling. The fourth row reveals results from 16× random sampling. Lastly, the seventh row depicts outcomes using 16× radial sampling.}
\label{fig4}
\end{figure}

\begin{figure}[!t]
\centerline{\includegraphics[width=1.01\columnwidth]{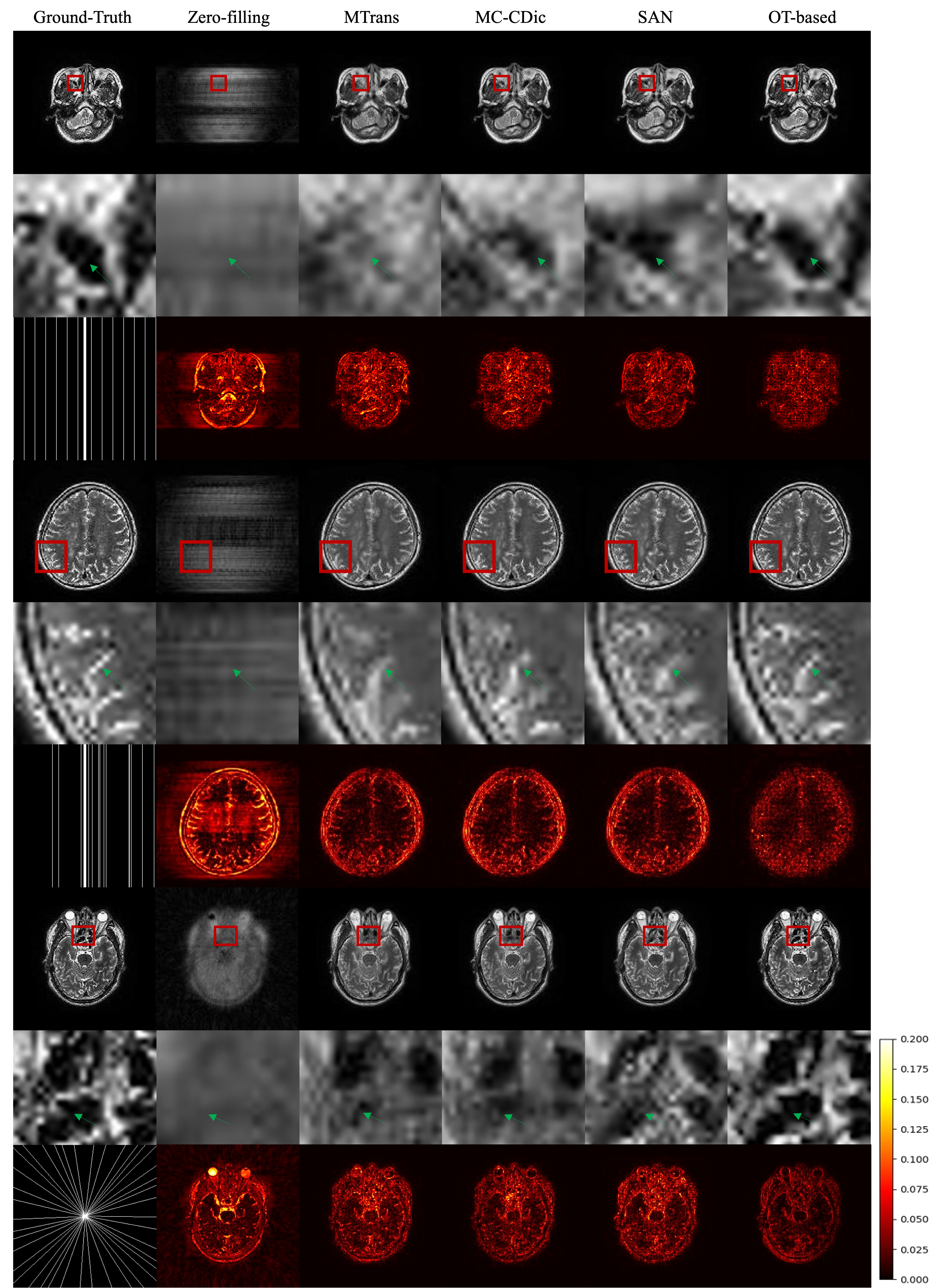}}
\caption{Comparison of the reconstruction results on the in-house dataset is shown. The first row shows outcomes from 16× equispaced sampling. The fourth row reveals results from 16× random sampling. Lastly, the seventh row depicts outcomes using 16× radial sampling.}
\label{fig5}
\end{figure}

The qualitative reconstruction results on the FastMRI dataset and the in-house dataset are depicted in Fig. \ref{fig4} and Fig. \ref{fig5}, providing insights into the performance of different methods. The first, fourth, and seventh rows showcase the reconstruction results of various methods at a sample ratio of $6.25\%$ using an equispaced pattern, a random pattern, and a radial pattern, respectively. The second, fifth, and eighth rows display the magnified detail images of these reconstructions. The third, sixth, and ninth rows present the corresponding error maps, which are obtained by subtracting the reconstructed T2-weighted images (T2WIs) from the real images. In the error map, regions appearing yellow indicate higher error magnitudes, while more pronounced structures indicate worse restoration quality.

Analyzing the results, we observe that the MTrans method exhibits noticeable aliasing artifacts and significant loss of anatomical structures. On the other hand, the SAN method, as a representative of cross-modal reconstruction, provides reconstructions with relatively clearer details. However, the OT-based method surpasses both methods by achieving further improvements. Specifically, at a sample ratio of $6.25\%$ using a random pattern, the OT-based method demonstrates the lowest reconstruction error and preserves anatomical details significantly.

\subsection{Ablation Study}
\label{discussion}

In this Section, we analyze the complementary nature of tasks, highlighting the importance of integrating multiple modalities for enhanced performance. Then, we demonstrate the essentiality of spatial and modal decomposition in the cross-modal synthesis process. Furthermore, We then delve into the significance of alternating iteration modules for image spatial alignment, emphasizing the necessity of addressing spatial misalignment to improve the synthesis process. Lastly, we provide a comprehensive analysis of the computational complexity of the proposed method.

\subsubsection{Complementary of Two Tasks}
\label{csp}

To assess the impact of the Cross-Modal Synthesis (CMS) process, a series of experiments was conducted on both FastMRI data and in-house data using different sampling patterns with random masks. The primary focus of these experiments was to evaluate the performance of the reconstruction process. Specifically, we compared the reconstructed images only from the under-sampled data with the fully-sampled ground truth images in a setting referred to as `Without CMS'. To measure the accuracy and quality of the reconstructed images, similar quantitative metrics were employed. The obtained results, presented in Table \ref{tb3}, illustrate that the reconstruction process effectively recovers missing details and enhances the overall fidelity of the reconstructed images, thanks to the integration of the cross-modal synthesis process.

\begin{table*}
\caption{Ablation study on the impact of cross-modal synthesis process. The experiments are carried out on two datasets using \textbf{random} masks, and the best result is indicated in bold.}
    \begin{center}
        \begin{tabular}{ccccccc}
            \toprule
            \multirow{2}*{Methods}  & \multicolumn{3}{c}{Sampling Ratio of $6.25\%$ on FastMRI Dataset}  & \multicolumn{3}{c}{Sampling Ratio of $6.25\%$ on In-House Dataset} \\
            \cline{2-7}
            & PSNR & SSIM & NMSE & PSNR & SSIM & NMSE\\
            \hline
             Without CMS	&33.48$\pm$1.63&0.9400$\pm$0.0200 &0.0189$\pm$0.0009
             &25.78$\pm$1.23&0.9431$\pm$0.0003 &0.1789$\pm$0.0103 \\
             With CMS &\textbf{36.30}$\pm$\textbf{1.27}&\textbf{0.9512}$\pm$\textbf{0.0061}&\textbf{0.0089}$\pm$\textbf{0.0004}
             &\textbf{27.56}$\pm$\textbf{1.05} &\textbf{0.9653}$\pm$\textbf{0.0033}&\textbf{0.1269}$\pm$\textbf{0.0021}\\
            \bottomrule
        \end{tabular}
    \end{center}\label{tb3}
\end{table*}

To demonstrate the impact of the information provided by low-quality images, this study compares the proposed method in this paper with MT-NET. The MT-NET method only utilizes information from the auxiliary modality for reconstruction. The paired dataset used for evaluation is $100\%$. The results on the FastMRI data and in-house data datasets are presented in Table \ref{tb4}. From the experimental results, it can be observed that even when reconstructing with low-quality images obtained at a sampling rate of $6.25\%$, our method still outperforms MT-NET. This indicates the need for simultaneous acquisition of low-quality to reconstruct high-quality target images. In the future, we will further improve the reconstruction results at even lower sampling rates.

\begin{table*}
\caption{Ablation study on the influence of information provided by low-quality target images. The first row presents the results of the MT-NET method on both the FastMRI dataset and our in-house dataset. This method does not utilize the information from low-quality T2WIs. The second and third rows display the results of the proposed method at a $6.25\%$ sampling rate.}
    \begin{center}
        \begin{tabular}{cccccccc}
            \toprule
              \multirow{2}*{Methods} & \multirow{2}*{Masks} & \multicolumn{3}{c}{FastMRI Dataset}  & \multicolumn{3}{c}{ In-House Dataset } \\
            \cline{3-8}
            & & PSNR & SSIM & NMSE & PSNR & SSIM & NMSE\\
            \hline
            MT-NET \cite{113} &--------& 30.31$\pm$2.13&0.9400$\pm$0.0200 &0.0189$\pm$0.0009
             &26.31$\pm$1.47&0.9511$\pm$0.0120 &0.1721$\pm$0.0211 \\
            \hline
            \multirow{2}*{OT-Based}
            &Random& 36.30$\pm$1.27 & 0.9512$\pm$0.0061 & 0.0089$\pm$0.0004 & 27.56$\pm$1.05 & 0.9653$\pm$0.0033 & 0.1269$\pm$0.0021 \\
            &Equispaced& 35.45$\pm$1.42 & 0.9573$\pm$0.0097 & 0.0143$\pm$0.0087 & 27.39$\pm$1.27 & 0.9652$\pm$0.0049 & 0.1433$\pm$0.0014 \\
            \bottomrule
        \end{tabular}
    \end{center}
    \label{tb4}
\end{table*}

\subsubsection{Spatial and Modal Decomposition in CMS Process}
\label{DS}

\begin{figure}[!htbp]
\centerline{\includegraphics[width=\columnwidth]{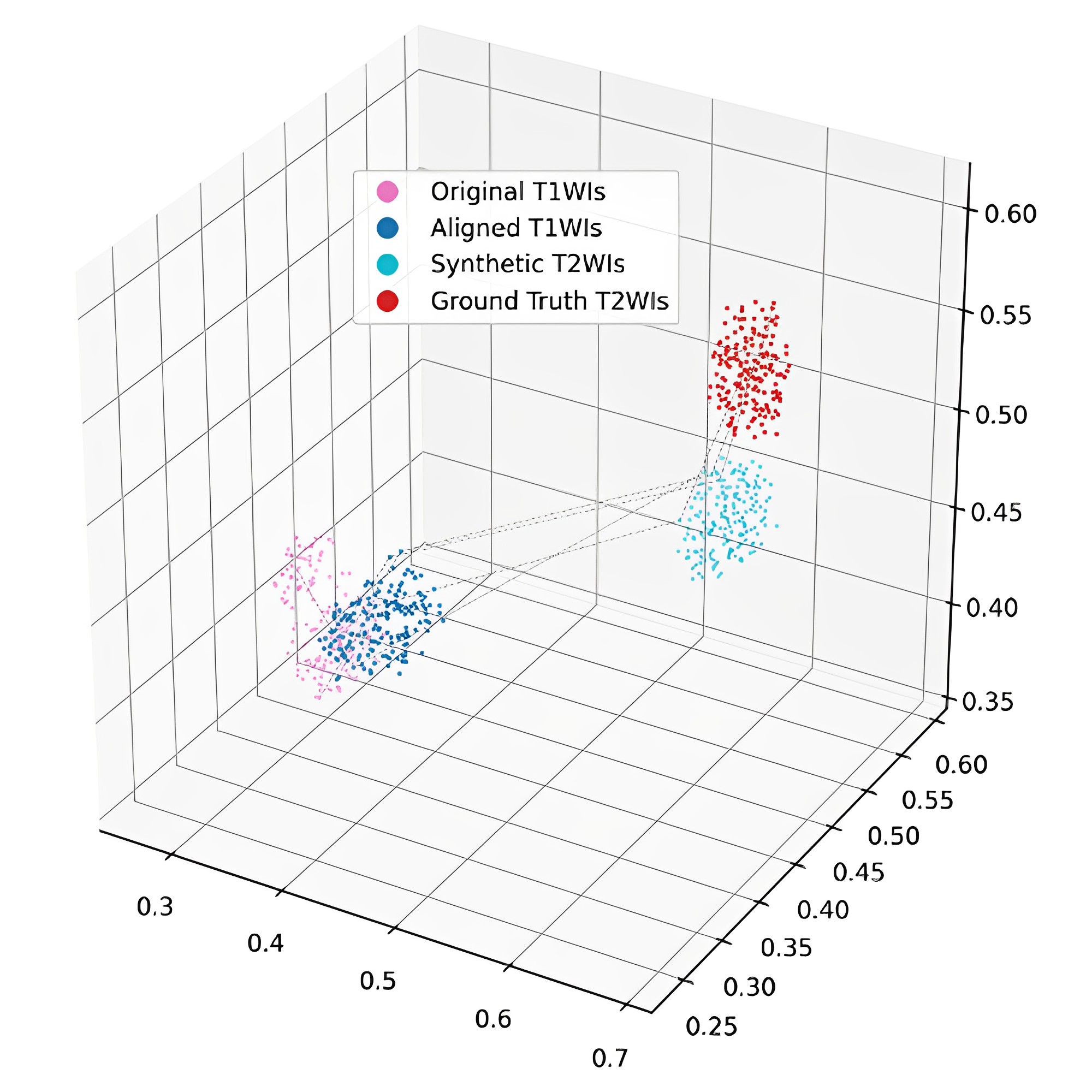}}
\caption{UMAP 3D Projection Analysis of MRI Image Distributions. The UMAP visualization shows the distribution patterns within a collection of 169 3D brain images from the FastMRI dataset, highlighting the distinctions between Original T1WIs, Aligned T1WIs, Synthetic T2WIs, and Ground Truth T2WIs.}
\label{fig15}
\end{figure}

To verify the necessity of eliminating spatial differences among target images and auxiliary images, we remove the effect of the Image Spatial Alignment (ISA) module. As a result, the original under-sampled target images and the fully-sampled auxiliary images directly are entered into the cross-modality synthesis network for high-quality target images. We define this setting as `Without ISA' and define the proposed method as `With ISA'. In the `Without ISA' setting, we directly input the misalignment images into the cross-modal synthesis network. From Table \ref{tb5}, the value of PSNR is improved from 33.99 dB to 35.45 dB, the value of SSIM from 0.9341 to 0.9573 and the value of NMSE is reduced from 0.0217 to 0.0143 on the FastMRI dataset. Notably, it is necessary to pay attention to the misalignment. Furthermore, the reconstruction performance will be advanced by integrating the OT-Based method. Further, we employed the UMAP technique for 3D projection of the data to assess the impact of the ISA module. In Fig. \ref{fig15}, the UMAP visualization of 169 3D brain images from the FastMRI dataset is presented, revealing the distribution characteristics among Original T1WIs, Aligned T1WIs, Synthetic T2WIs, and Ground Truth T2WIs. The findings demonstrated that Aligned T1WIs are closer to the Ground Truth T2WIs, validating the effectiveness of the ISA module. Similarly, the close grouping of Synthetic T2WIs with Ground Truth T2WIs showcased the high fidelity of the synthesis process.

\begin{table*}
\caption{Ablation study on the essentiality of spatial and modal decomposition in cross-modal synthesis process. The experiments are carried out on two datasets using \textbf{Equispaced} masks, and the best result is indicated in bold.}
    \begin{center}
        \begin{tabular}{ccccccc}
            \toprule
            \multirow{2}*{Methods}  & \multicolumn{3}{c}{Sampling Ratio of $6.25\%$ on FastMRI Dataset} & \multicolumn{3}{c}{Sampling Ratio of $6.25\%$ on In-House Dataset} \\
            \cline{2-7}
            & PSNR & SSIM & NMSE & PSNR & SSIM & NMSE \\
            \hline
             Without ISA	&33.99$\pm$1.74	&0.9341$\pm$0.0107 & 0.0217$\pm$0.0071 &26.89$\pm$1.39	&0.9423$\pm$0.0109 & 0.1799$\pm$0.0082\\
             With ISA &\textbf{35.45}$\pm$\textbf{1.42}& \textbf{0.9573}$\pm$\textbf{0.0097}&\textbf{0.0143}$\pm$\textbf{0.0087}            &\textbf{27.39}$\pm$\textbf{1.27}&\textbf{0.9652}$\pm$\textbf{0.0049} &\textbf{0.1433}$\pm$\textbf{0.0014}\\
            \bottomrule
        \end{tabular}
    \end{center}\label{tb5}
\end{table*}

\subsubsection{Alternating Iteration Strategy for Image Spatial Alignment}
\label{ISA}

\begin{figure}[!htbp]
\centerline{\includegraphics[width=\columnwidth]{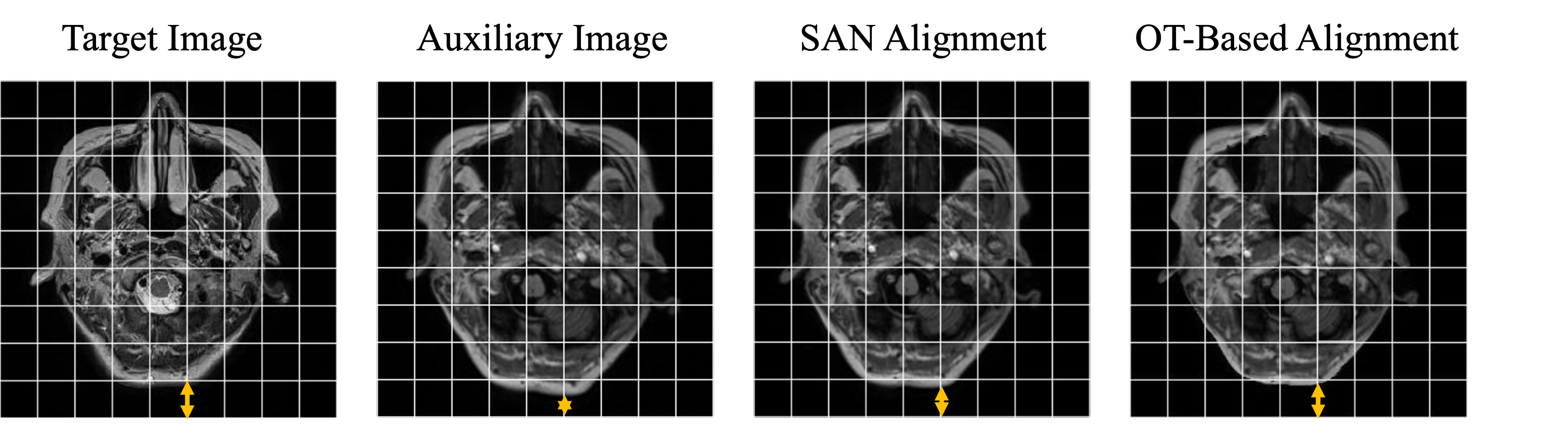}}
\caption{A subtle spatial misalignment between the auxiliary (T1-weighted) image and the target (T2-weighted) image is noticeable in this figure. To address this concern, the approach SAN proposed in \cite{8} utilizes a spatial alignment technique to produce an aligned-auxiliary image. `SAN Alignment' refers to the T1-weighted image warped using the SAN method proposed in \cite{8}, whereas the `OT-Based Alignment' is generated using our own method.}
\label{fig6}
\end{figure}

In this section, we show our superior alignment accuracy and improved preservation of structural details. As depicted in Fig. \ref{fig6}, upon the length of the yellow bidirectional, we can observe that the proposed method achieves precise spatial alignment during the cross-modal synthesis process. The integration of alignment with the reconstruction process leads to enhanced alignment accuracy, improved preservation of structural details, and overall superior performance. This demonstrates the effectiveness of our approach in achieving accurate alignment and producing visually coherent synthesized images.

\subsubsection{Computational Complexity}
\label{cc}

Computational complexity is a crucial measure for evaluating algorithm performance. Table \ref{tb6} presents three key complexity metrics, including the number of parameters, the amount of multiply-accumulates (MACs), and memory requirements. MACs are measured using the ptflops tool, while memory footprint is computed on a Nvidia GeForce RTX GPU. In this study, both the proposed method and SAN consider the issue of spatial misalignment, allowing for a fair comparison of their computational complexities.

Comparing the computational complexities of the proposed method with SAN, the OT-Based method exhibits superior results with fewer parameters, lower MACs, and reduced memory requirements. For instance, the proposed method only requires 27.86 M parameters, whereas SAN necessitates 47.23 M parameters. The proposed method significantly reduces the parameter count compared to SAN. Notably, our method also achieves a notable reduction in memory requirements, reducing it from 352.21 MiB to 224.97 MiB. These results clearly demonstrate the effectiveness of the proposed algorithm.

\begin{table}
\caption{Computational complexity comparison on the FastMRI datdaset.}
    \begin{center}
        \begin{tabular}{cccc}
            \toprule
            Methods & Parameters (M) & MACs (G) & Memory(MiB)\\
            \hline
            SAN \cite{8} & 47.23 & 174.71 & 352.21\\
            OT-Based	&\textbf{27.86} & \textbf{104.01} & \textbf{224.97}\\
            \bottomrule
        \end{tabular}\label{tb6}
    \end{center}
\end{table}

\section{CONCLUSION}
\label{sec:DISCUSSION AND CONCLUSION}
In this paper, our focus is on accelerating reconstruction of T2WIs. To achieve it, we employ a reconstruction network that takes the under-sampled T2WI as input and produces a high-quality reconstructed T2WI. In order to leverage the complementary information across different modalities, we introduce a cross-modal synthesis process to generate synthetic T2WI. We prove that the cross-modal synthesis task and the reconstruction task mutually support each other by minimizing the difference between the reconstructed T2WI and the synthetic one in the $T2$ manifold. Moreover, to eliminate the negative transfer of subtle misalignment between paired images from the T1-weighted modality and the T2-weighted modality, we decompose the cross-modal synthesis process into two OT processes. The one OT process focuses on spatial alignment mapping of the T1-weighted modality, while the other OT process synthesizes the mapping from the T1-weighted modality to the T2-weighted modality. This decomposition enhances the interpretability of the cross-modal synthesis process. Finally, experimental evaluations on both the FastMRI dataset and an in-house whole brain MRI dataset confirm the effectiveness of the proposed method.

\printbibliography
\end{document}